\documentclass{article}

\usepackage{PRIMEarxiv}

\usepackage[utf8]{inputenc} 
\usepackage[T1]{fontenc}    
\usepackage{hyperref}       
\usepackage{url}            
\usepackage{booktabs}       
\usepackage{amsfonts}       
\usepackage{nicefrac}       
\usepackage{microtype}      
\usepackage{lipsum}
\usepackage{fancyhdr}       
\usepackage{graphicx}       
\graphicspath{{media/}}     

\title{Rewriting TTS Inference Economics: Lightning V2 on Tenstorrent Achieves 4× Lower Cost Than NVIDIA L40S
}

\author{
  Ranjith M S \\
  Senior AI Inference Performance Engineer \\
  Smallest AI \\
  \texttt{ranjith@smallest.ai} \\
   \And
  Akshat Mandloi \\
  CTO \\
  Smallest AI \\
  \texttt{akshat@smallest.ai} \\
  \And
  Sudarshan Kamath \\
  CEO \\
  Smallest AI \\
  \texttt{sudarshan@smallest.ai} \\
}

\begin{document}
\maketitle

\begin{abstract}
Text-to-Speech (TTS) models are significantly more numerically fragile than Large Language Models (LLMs) due to their continuous waveform generation and perceptual sensitivity to small numerical perturbations. While aggressive precision reduction techniques such as BlockFloat8 (BFP8) and low-fidelity (LoFi) compute have been widely adopted in language models, applying similar strategies to TTS systems often results in audible artifacts, phase instability, and spectral distortion.

In this work, we present Lightning V2, a production-grade TTS model co-optimized for Tenstorrent hardware. Through precision-aware architectural design and hardware–software co-optimization, we achieve over 95\% LoFi computational fidelity and more than 80\% BlockFloat8 deployment without measurable degradation in audio quality. Leveraging Tenstorrent’s Network-on-Chip (NoC), distributed SRAM, and deterministic execution model, we reduce memory movement and redundant weight fetches, enabling efficient low-precision inference.

Compared to an NVIDIA L40S baseline, Lightning V2 achieves approximately 4× lower on-prem accelerator cost at equivalent concurrency, while maintaining production audio fidelity. Our results demonstrate that precision co-design, combined with hardware-aware optimization, can fundamentally reshape the economics of real-time speech inference.
\end{abstract}


\section{Introduction}

\subsection{Motivation}

Text-to-Speech systems have rapidly evolved from research prototypes to production-critical infrastructure powering voice assistants \cite{hoy2018alexa}, accessibility tools, conversational agents, and real-time communication systems \cite{guo2024flyttsfastlightweighthighquality} \cite{wu2025clearcontinuouslatentautoregressive} \cite{li2024cmttsenhancingrealtime}. As adoption increases, inference cost -- rather than training cost -- becomes the dominant economic factor, particularly for latency-sensitive and on-prem deployments \cite{10363447} \cite{Luccioni_2024}.

Recent advances in Large Language Models (LLMs) have demonstrated that aggressive precision reduction techniques such as FP8 \cite{micikevicius2022fp8}, BlockFloat8 (BFP8) \cite{wikipediaBlockFloatingPoint}, and low-fidelity (LoFi) \cite{tenstorrentMatrixEngine} compute can significantly reduce inference cost without materially impacting output quality. These techniques reduce memory bandwidth requirements, improve compute efficiency, and enable higher concurrency per accelerator. However, directly transferring these optimizations to TTS systems has proven challenging.

Unlike LLMs, TTS models generate continuous waveforms through multi-stage pipelines involving diffusion-based acoustic modeling \cite{popov2021gradttsdiffusionprobabilisticmodel} \cite{ju2024naturalspeech3zeroshotspeech} and neural vocoders\cite{sun2023aisynthesizedvoicedetectionusing}. Small numerical perturbations can accumulate across timesteps, alter harmonic structure, and introduce perceptually noticeable artifacts \cite{kim2024quantization}. As a result, TTS inference remains heavily reliant on higher precision formats, limiting achievable cost reductions.

Reducing inference cost for TTS without compromising perceptual quality remains an open systems challenge.

\subsection{Problem Statement}

Aggressive numerical optimization in TTS presents two fundamental challenges. First, speech generation is numerically fragile: minor deviations in intermediate activations can manifest as phase distortion, pitch instability, metallic ringing, or temporal artifacts in the final waveform. Unlike token-based LLM models, TTS systems lack natural reset boundaries; numerical error directly influences continuous signal reconstruction.

Second, memory movement dominates inference cost in modern accelerators. Conventional GPU architectures rely heavily on global memory round-trips between layers and across execution units. While techniques such as batching and high-bandwidth memory mitigate throughput bottlenecks, single-sample or low-batch real-time TTS inference remains constrained by latency and memory traffic.

The central question we address is:

\begin{quote}
Can we aggressively reduce numerical precision and compute fidelity in a production-grade TTS system while preserving audio quality, and can hardware--software co-design fundamentally reduce inference cost?
\end{quote}

\subsection{Contributions}

In this work, we present Lightning V2, a diffusion-based Text-to-Speech model co-optimized for Tenstorrent hardware. Our key contributions are:

\begin{itemize}
    \item \textbf{Precision-Aware TTS Optimization:} We demonstrate that over 95\% of layers can operate in LoFi computational fidelity while preserving perceptual audio quality.
    
    \item \textbf{High BlockFloat8 Adoption:} We achieve more than 80\% BlockFloat8 deployment across the model, resulting in approximately 2$\times$ model size reduction and significant memory transfer savings.
    
    \item \textbf{Hardware--Software Co-Design:} By leveraging Tenstorrent's Network-on-Chip, distributed SRAM, multicast weight delivery, and deterministic execution model, we reduce redundant DRAM traffic and improve effective throughput.
    
    \item \textbf{Cost-Equivalent Concurrency Gains:} At comparable utilization levels (550 simultanious TTS requests), our system achieves approximately 4$\times$ lower accelerator cost compared to an NVIDIA L40S baseline.
    
    \item \textbf{Empirical Study of Numerical Fragility:} We provide practical insights into the limitations of conventional similarity metrics (e.g., PCC) for TTS optimization and highlight the gap between tensor-level similarity and perceptual fidelity.
\end{itemize}

Together, these results demonstrate that precision co-design combined with hardware-aware optimization can significantly reshape the economics of real-time speech inference.

\section{Background}

\subsection{Numerical Optimization in Large Language Models}

Over the past several years, large language models have undergone a steady reduction in numerical precision during both training and inference. Early large-scale models were predominantly trained and deployed in FP32\cite{30711}, later transitioning to FP16\cite{micikevicius2018mixedprecisiontraining} to improve memory efficiency and throughput. BF16\cite{kalamkar2019studybfloat16deeplearning} subsequently became widely adopted due to its larger exponent range relative to FP16, offering improved numerical stability with comparable storage cost.

More recently, FP8 formats have emerged as viable alternatives for inference and, in some cases, training. FP8 reduces memory bandwidth requirements and increases arithmetic throughput, particularly on hardware architectures with native FP8 support. In parallel, low-fidelity (LoFi) approaches trade numerical precision—primarily through reduced mantissa width—for higher compute efficiency. In contrast, Block Floating Point formats preserve similar numerical precision while reducing memory and bandwidth costs by sharing exponents across groups of values.

These techniques have proven effective in LLMs due to several properties of language modeling workloads:
\begin{itemize}
    \item Token-level objectives are relatively tolerant to small logit perturbations.
    \item Errors introduced at intermediate layers often remain bounded after softmax normalization.
    \item Autoregressive decoding mitigates local deviations through subsequent context updates.
\end{itemize}

As a result, aggressive precision reduction—down to FP8 or block floating-point formats—often incurs minimal degradation in perplexity or downstream task metrics.

\subsection{Diffusion-Based Text-to-Speech Architecture}

Modern high-quality text-to-speech systems are often described as a two-stage pipeline consisting of:

\begin{enumerate}
\item \textbf{Acoustic Model:} A generative model (frequently diffusion-based or transformer-based) that maps linguistic or phoneme representations to intermediate acoustic features.
\item \textbf{Neural Vocoder:} A waveform generator that converts these features into time-domain audio samples.
\end{enumerate}

However, recent systems increasingly depart from this strict decomposition, instead operating directly in learned audio latent spaces or generating waveform representations end-to-end, eliminating explicit spectrogram intermediates.

Diffusion-based approaches iteratively refine a noisy latent representation over multiple denoising steps. Unlike autoregressive language models, these models operate over continuous signal representations and optimize objectives tied to reconstruction fidelity in acoustic or latent feature spaces.

Audio generation introduces strong sensitivity to small numerical deviations. Perturbations in intermediate representations—whether spectrograms or learned latent features—can propagate into perceptible artifacts in the final waveform. This sensitivity is further amplified during waveform synthesis, where phase and harmonic structure must remain coherent across thousands of output samples.

Consequently, TTS systems often exhibit tighter numerical stability requirements than token-based language models.

\subsection{Tenstorrent Architecture}

Tenstorrent accelerators employ a distributed dataflow architecture characterized by:

\paragraph{Network-on-Chip (NoC).}
Cores communicate via a packet-based network-on-chip, enabling explicit data movement between compute units. This design reduces reliance on centralized memory hierarchies and supports fine-grained scheduling of tensor tiles.

\paragraph{Distributed On-Chip SRAM.}
Each compute core is paired with local SRAM, allowing high-bandwidth access to frequently reused data. Compared to architectures that rely heavily on external DRAM, this reduces memory traffic and associated energy cost.

\paragraph{1:1 Thread-to-Core Mapping.}
Workloads are explicitly mapped such that one software thread corresponds to one hardware core. This mapping enables predictable execution patterns and reduces scheduling overhead.

These architectural characteristics influence how low-precision arithmetic and block floating-point formats can be exploited during inference, particularly for workloads with high data reuse and structured tensor movement patterns.
\subsubsection{Memory Hierarchy and Data Movement Bandwidth}

Efficient data movement is critical for achieving high performance in dataflow-oriented accelerators. Table~\ref{tab:memory_bandwidth} summarizes the effective bandwidth across different memory and communication patterns in the system.

On-chip SRAM provides the highest bandwidth when data is locally reused or shared across nearby compute units. As communication distance increases (e.g., multi-hop gather/scatter), effective bandwidth decreases due to network contention and routing overhead. Multicast patterns enable efficient distribution of shared weights, but still operate below peak local bandwidth.

In contrast, off-chip DRAM access is significantly more bandwidth-constrained, reinforcing the importance of minimizing DRAM round-trips through SRAM-resident tiling and reuse. Network-based communication (e.g., Ethernet) provides flexibility for inter-system scaling but is not suitable for latency-sensitive inner-loop execution.

\begin{table}[h]
\centering
\begin{tabular}{lll}
\hline
\textbf{Memory \& I/O} & \textbf{Data Movement Pattern} & \textbf{Bandwidth} \\
\hline
SRAM & Local / Shared & 94 TB/s \\
SRAM & Neighbor (Halo) & 47 TB/s \\
SRAM & Row / Column / Mesh Multicast & 24 TB/s \\
SRAM & Gather / Scatter (3 hops) & 16 TB/s \\
SRAM & Gather / Scatter (10 hops) & 5 TB/s \\
DRAM & Row Access & 512 GB/s \\
Ethernet & Column Communication & 1 TB/s \\
\hline
\end{tabular}
\caption{Effective bandwidth across memory hierarchy and data movement patterns in P150. \cite{vasiljevic2024blackhole}}
\label{tab:memory_bandwidth}
\end{table}


\subsubsection{Tensix Core Microarchitecture and Execution Model}

\begin{figure}[h]
\centering
\includegraphics[width=\linewidth]{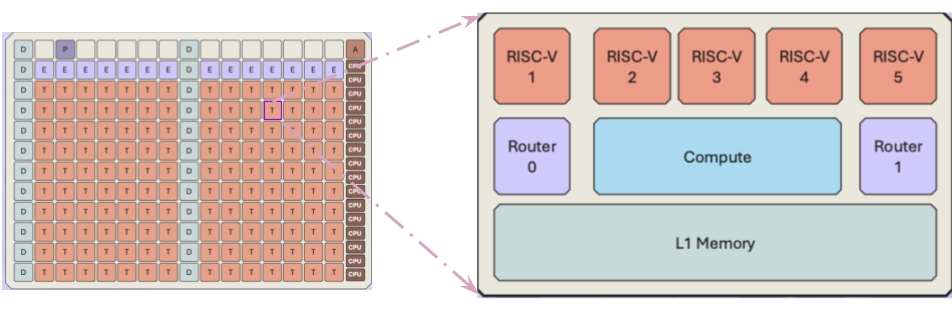}
\caption{Spatial layout of Tensix cores and NoC connectivity.}
\label{fig:tensix_grid}
\end{figure}

Figure~\ref{fig:tensix_grid} illustrate the spatial organization of Tensix cores and the internal microarchitecture of a single core. Each Tensix core follows a dataflow execution model, where computation and data movement are explicitly orchestrated in software.

\paragraph{Five-Stage Asynchronous Pipeline.}
Each core executes five independent pipelines concurrently:

\begin{itemize}
    \item \textbf{Reader (RISC-V 1):} Fetches tiles from DRAM or remote cores via the NoC into on-chip SRAM.
    \item \textbf{Unpacker (RISC-V 2):} Transfers tiles from SRAM circular buffers to compute engines.
    \item \textbf{Compute (RISC-V 3,4):} Executes matrix and vector operations on the math and SIMD engines.
    \item \textbf{Writer (RISC-V 5):} Writes computed tiles back to DRAM or transmits them across the NoC.
\end{itemize}

These stages operate \emph{asynchronously}, allowing overlap of data movement and computation. Synchronization is achieved through producer–consumer counters rather than global barriers.

\paragraph{On-Chip SRAM and Circular Buffers.}
Each core is equipped with 1.5\,MB of local SRAM, explicitly managed by software. Data is organized into circular buffers (CBs), which decouple pipeline stages:

\begin{itemize}
    \item Reader populates input CBs
    \item Compute consumes input CBs and produces output CBs
    \item Writer drains output CBs
\end{itemize}

This design eliminates the need for hardware-managed caches and enables deterministic data reuse and scheduling.

\paragraph{Compute Engines.}
Each core contains:

\begin{itemize}
    \item \textbf{Math Engine:} Matrix multiply-accumulate (MAC) arrays operating on tiled inputs (e.g., $32 \times 32$ tiles)
    \item \textbf{SIMD Engine:} Supports FP32, FP16, BF16, FP8, and INT8 operations
\end{itemize}

The separation of data movement and compute allows sustained utilization even under irregular workloads.

\paragraph{Dataflow Execution.}
A typical execution proceeds as follows:

\begin{enumerate}
    \item Tiles are fetched from DRAM or neighboring cores via the NoC
    \item Tiles are placed into SRAM circular buffers
    \item Compute engines operate on tiles as they become available
    \item Results are written back to SRAM buffers
    \item Output tiles are transmitted to DRAM or downstream cores
\end{enumerate}

Because each stage operates independently, multiple tiles can be in-flight simultaneously, forming a streaming pipeline across the core.

\paragraph{Contrast with GPU Execution Models.}
This execution model differs fundamentally from conventional GPU programming:

\begin{itemize}
    \item \textbf{CUDA:} A single kernel is launched, and hardware dynamically schedules warps. Memory hierarchy (L1/L2 caches) is managed implicitly.
    \item \textbf{Tensix:} Separate reader, compute, and writer kernels are explicitly defined. Data placement in SRAM and movement across cores must be manually orchestrated.
\end{itemize}

This explicit dataflow model enables fine-grained control over memory movement and eliminates redundant data transfers, but requires careful co-design between software and hardware.

\paragraph{Implications for TTS Inference.}
The decoupled pipeline and explicit SRAM management are particularly well-suited for TTS workloads, where intermediate activations can be retained on-chip and reused across timesteps. Combined with NoC multicast, this allows efficient distribution of shared weights while minimizing DRAM bandwidth pressure.
\section{Numerical Fragility and Systems Methodology}

Text-to-speech inference differs fundamentally from token-based language modeling due to its operation in continuous signal space. In this section, we characterize the numerical fragility of diffusion-based TTS systems and describe the hardware–software strategies used to mitigate degradation under reduced-precision execution.

\subsection{Continuous Signal Sensitivity}

Unlike LLMs that operate over discrete token probabilities, TTS systems generate continuous-valued acoustic representations and ultimately time-domain waveforms. Small perturbations in intermediate activations directly modify frequency amplitudes, phase relationships, and harmonic structure.

In practice, we observed that minor rounding-induced deviations—while numerically small—can produce perceptible distortions in the synthesized waveform. These include:
\begin{itemize}
    \item High-frequency ringing artifacts
    \item Pitch instability
    \item Temporal smearing across frames
\end{itemize}

Such artifacts are not easily captured by conventional tensor-level similarity metrics, highlighting a fundamental mismatch between numerical error and perceptual quality.

\subsection{Diffusion Error Accumulation}

Diffusion-based acoustic models refine representations iteratively across multiple denoising steps. Each step depends on the previous latent estimate. Consequently, small rounding errors introduced at early timesteps propagate and may compound over the denoising trajectory.

Unlike autoregressive LLM decoding, which resets normalization boundaries at each token prediction, diffusion operates over a persistent latent state. Reduced-precision perturbations therefore influence the entire denoising path.

We observed cases where individual layers exhibited near-perfect intermediate correlation with higher-precision baselines, yet cumulative error across diffusion steps resulted in audible degradation.

\subsection{Dynamic Range Sensitivity}

Speech signals exhibit wide dynamic range across time and frequency. Low-energy regions (e.g., fricatives or silence transitions) are particularly sensitive to quantization error, as relative perturbation becomes large compared to signal magnitude.

Reduced mantissa precision and shared-exponent formats (e.g., block floating point) may introduce non-uniform distortion depending on local signal statistics. This requires selective application of low-precision arithmetic rather than uniform global quantization.

\subsection{Metric Misalignment: A PCC Case Study}

One of the most surprising observations during experimentation concerned the reliability of Pearson Correlation Coefficient (PCC), which is commonly treated as a gold-standard numerical similarity metric.

When comparing the PyTorch model executed on an NVIDIA L40S GPU against CPU execution (AMD EPYC 7352 24-Core Processor), the end-to-end PCC between outputs was approximately 0.72, despite identical inputs and equivalent model weights. Numerically, such a PCC value would typically be interpreted as significant deviation or mismatch. However, the generated audio waveforms were perceptually indistinguishable and of high quality.

This discrepancy complicated the porting process to Tenstorrent hardware, as no single numerical metric reliably captured correctness.

In a separate debugging instance, a particular layer exhibited:
\begin{itemize}
    \item Extremely high PCC (rounded to 1.0)
    \item Very small relative error
\end{itemize}

By conventional numerical standards, the layer appeared correct. Yet, enabling reduced-precision execution in that layer consistently produced audible degradation in the final waveform.

Pinpointing this issue required over a month of systematic investigation. Initial debugging efforts focused on layers exhibiting lower PCC values, under the assumption that these would be responsible for output divergence. Ironically, the problematic layer appeared numerically “perfect” by standard tensor similarity metrics.

This experience highlights a critical insight:

\begin{quote}
Traditional numerical similarity metrics such as PCC or relative error are not reliable indicators of perceptual audio quality in TTS systems.
\end{quote}

End-to-end perceptual validation is therefore necessary when evaluating reduced-precision deployments for continuous signal generation.

\subsection{LoFi Computational Fidelity}

To balance efficiency and perceptual stability, we introduced controlled reductions in computational fidelity. LoFi execution reduces arithmetic precision for selected operations while retaining sufficient dynamic range to avoid catastrophic drift.

Rather than applying uniform quantization, we defined discrete fidelity levels corresponding to varying mantissa precision and accumulation strategies. Lower levels increase throughput but require empirical validation against perceptual degradation.

Only operations empirically determined to be numerically tolerant were executed under reduced fidelity.

\subsection{BlockFloat8 Deployment Strategy}

Block Floating Point was deployed selectively across the model. BFP8 shares an exponent across blocks of values, improving compute density while preserving exponent range.

However, not all layers exhibited sufficient tolerance to block-wise exponent sharing. Layers exhibiting high dynamic range or diffusion-state sensitivity retained higher precision formats.

Layer selection was guided by:
\begin{itemize}
    \item End-to-end perceptual evaluation
    \item Sensitivity to diffusion-step perturbations
    \item Dynamic range characteristics
\end{itemize}

This selective strategy prevented compounding instability while enabling substantial reduction in memory traffic.

\subsection{Custom Kernel Implementations}

Certain computational kernels exhibited performance bottlenecks or heightened numerical sensitivity under reduced precision.

We implemented custom kernels to:
\begin{itemize}
    \item Improve data locality
    \item Reduce redundant memory movement
\end{itemize}

These kernels were optimized to preserve numerical stability while exploiting hardware-level parallelism.

\subsection{Hardware Co-Design Strategy}

Performance gains were further achieved through explicit hardware–software co-design:

\paragraph{Multicast via Network-on-Chip.}
Frequently reused weights were multicast across compute cores using the on-chip network, reducing redundant DRAM fetches.

\paragraph{SRAM-Aware Tiling.}
Tensor tiles were structured to maximize reuse within local SRAM, minimizing global memory traffic.

\paragraph{DRAM Round-Trip Avoidance.}
Intermediate activations were kept on-chip whenever possible, avoiding unnecessary external memory transfers.

Together, these strategies reduced memory bandwidth pressure—one of the dominant cost contributors in inference workloads.

\section{Experimental Evaluation}

\subsection{Hardware Platforms}

We evaluate Lightning V2 inference across the following accelerators:

\begin{itemize}
    \item NVIDIA L40S GPU
    \item Tenstorrent P100
    \item Tenstorrent P150
\end{itemize}

All experiments were conducted under comparable software configurations, with identical model weights and inference workloads.

The P100 and P150 are expected to exhibit identical single-chip latency for this workload, as Lightning V2 does not utilize multi-chip execution or high-speed interconnect (QSFP). Reported single-chip latencies correspond to measured P150 results.






\subsection{Speech Quality and Semantic Fidelity}

We evaluate the impact of hardware-aware inference on speech quality and semantic consistency by comparing outputs generated on NVIDIA L40s and Tenstorrent P150. We report DNSMOS as a perceptual quality metric and Word Error Rate (WER) as a measure of semantic fidelity.

\begin{table}[h]
\centering
\caption{Comparison of NVIDIA and Tenstorrent TTS outputs}
\begin{tabular}{lcc}
\toprule
\textbf{Metric} & \textbf{NVIDIA} & \textbf{Tenstorrent} \\
\midrule
DNSMOS $\uparrow$ & 3.872 & 3.801 \\
\midrule
\multicolumn{3}{l}{\textbf{Relative Difference}} \\
$\Delta$ DNSMOS (P150 - L40s) & \multicolumn{2}{c}{-0.071} \\
\midrule
WER (normalized) $\downarrow$ & \multicolumn{2}{c}{0.009} \\
\bottomrule
\end{tabular}
\end{table}

The results indicate that Tenstorrent inference closely preserves semantic content, with a normalized WER of 0.009, suggesting near-identical transcriptions between the two systems. This confirms that the underlying linguistic information is largely unaffected by the change in hardware and numerical precision.

In terms of perceptual quality, Tenstorrent exhibits a modest degradation, with a DNSMOS drop of 0.071 compared to NVIDIA. This difference is relatively small and falls within the range of minor perceptual variation, indicating that speech naturalness is largely retained despite reduced precision.

Overall, these results demonstrate that hardware-efficient inference can maintain semantic fidelity while incurring only a limited impact on perceptual quality, highlighting a favorable trade-off between efficiency and output quality.

\subsection{Cost and Concurrency Comparison}

\paragraph{Single-Device Comparison.}

We first compare per-device performance under steady-state inference.

\begin{table}[htbp]
\centering
\begin{tabular}{lcccc}
\hline
Hardware & Cost (USD) & Concurrency & Latency (ms) & Cost Gain \\
\hline
L40S & \$9000 & 3 & 300 & 1$\times$ \\
P150 & \$1400 & 1 & 250 & 2.6$\times$ \\
P100 & \$1000 & 1 & 250 & 3.6$\times$ \\
\hline
\end{tabular}
\caption{Single-Device Comparison of Latency, Concurrency, and Cost}
\label{tab:single_device_comparison}
\end{table}

The L40S sustains higher per-device concurrency; however, the Tenstorrent devices achieve lower per-request latency at significantly reduced hardware cost. When normalized by sustained request capacity, the cost per concurrent TTS stream is substantially lower on Tenstorrent hardware.

\medskip

\noindent
\textit{Note:} The reported single-device latency for P100 corresponds to measured P150 latency. Lightning V2 executes on a single chip without multi-chip parallelism or QSFP interconnect usage. Therefore, single-chip latency is expected to be equivalent between P100 and P150 for this workload.

\paragraph{Fleet-Level Extrapolation.}

\begin{figure}[h]
\centering
\includegraphics[width=0.8\linewidth]{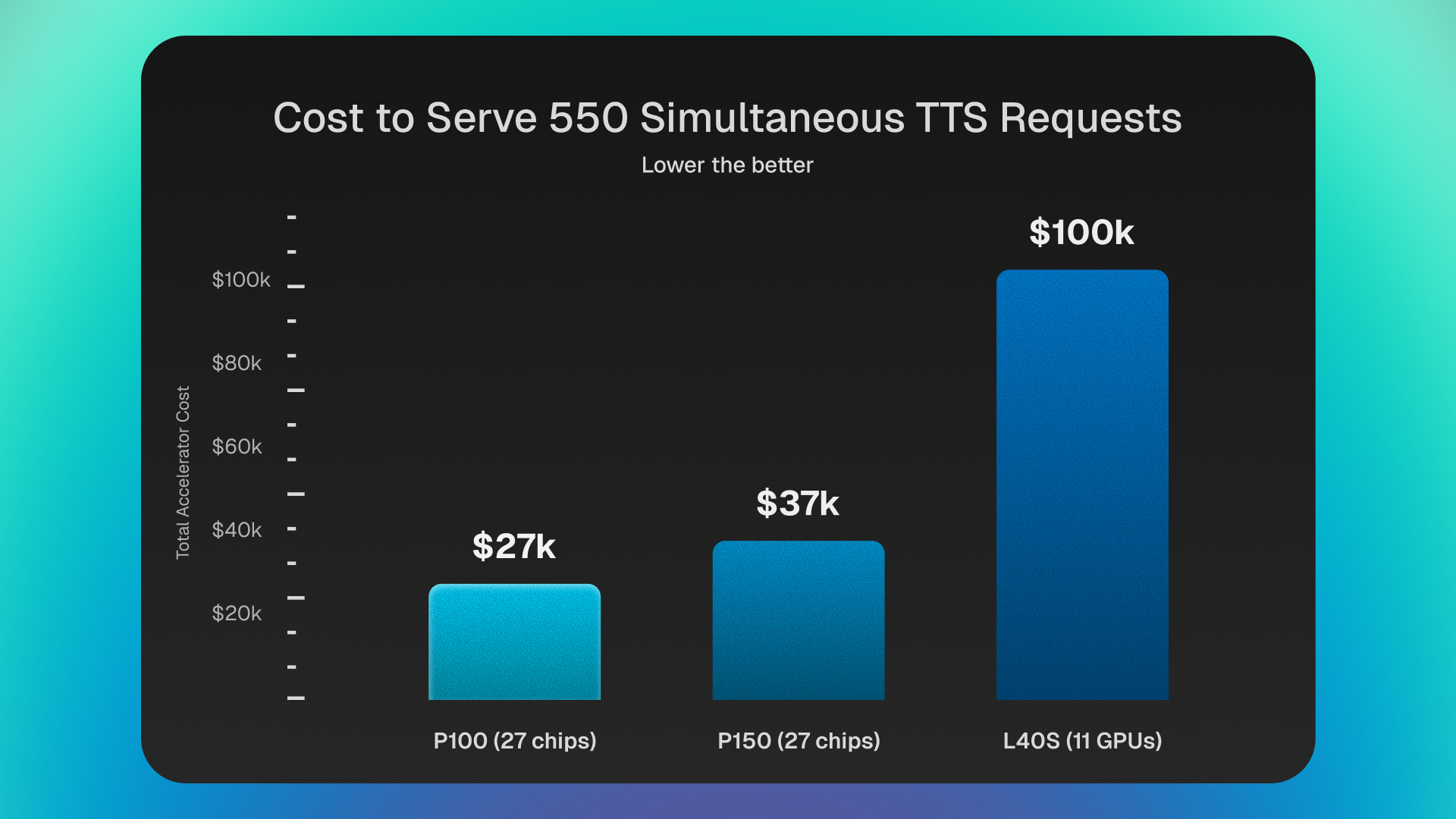}
\caption{Accelerator cost to sustain 550 5-second TTS requests, showing a 3--4$\times$ lower cost for Tenstorrent vs.\ NVIDIA.}
\label{fig:cost}
\end{figure}

\noindent

\textbf{Assumption (for simplicity):} Each response produces approximately 5 seconds of audio (this may vary slightly per request). From Table \ref{tab:single_device_comparison}, each L40S produces approximately $3.33 \times 3 \times 5 \approx 50$ seconds of audio per second, while a P150 produces $4 \times 5 = 20$ seconds of audio per second.

\medskip

\noindent
We consider a voice agent pipeline (AST $\rightarrow$ LLM $\rightarrow$ TTS) operating at 100\% hardware utilization, where the next input to the TTS stage is always ready before the current generation completes.

\medskip

\noindent
To sustain a workload equivalent to generating $550 \times 5 = 2750$ seconds of audio within a 5-second window—corresponding to a steady stream of 550 overlapping 5-second requests at full utilization:

\medskip

\noindent
A single NVIDIA GPU can process approximately 10 requests per second (based on the measured $\sim$300 ms for 3 responses), delivering $10 \times 5 = 50$ audio-seconds per second. Therefore, 11 GPUs collectively provide $\sim$550 audio-seconds per second, which is sufficient to sustain this workload.

Similarly, a single Tenstorrent P150 produces $\sim$20 audio-seconds per second, implying that $\sim$27 devices are required to sustain the target throughput of 550 audio-seconds per second.
\medskip

\noindent
This translates to the following infrastructure requirements:
\begin{itemize}
    \item 11 $\times$ NVIDIA L40S GPUs ($\sim$\$100{,}000 total accelerator cost)
    \item 27 $\times$ Tenstorrent P100 accelerators ($\sim$\$27{,}000 total accelerator cost)
    \item 27 $\times$ Tenstorrent P150 accelerators ($\sim$\$37{,}000 total accelerator cost)
\end{itemize}

\medskip

\noindent
This represents a $\sim$3--4$\times$ reduction in upfront accelerator cost to serve the same workload. The difference between $\sim$\$27K and $\sim$\$100K is not incremental—it is decisive. For many deployments, that delta alone determines whether on-prem inference is feasible.

\subsection{Throughput and Layer-Level Performance}

While the current results demonstrate substantial hardware leverage, several program configurations remain sub-optimal, leaving meaningful performance headroom.

To illustrate this potential, one production layer in Lightning V2 (approximately 6B MACs) currently exhibits:

\begin{itemize}
    \item $\sim$60 $\mu$s execution time on NVIDIA L40S
    \item $\sim$31 $\mu$s on Tenstorrent P150
\end{itemize}

This 2$\times$ latency improvement is achieved on a single layer without exhaustive global tuning. When normalized by accelerator cost, the effective performance-per-dollar improvement for this layer exceeds an order of magnitude.

The magnitude of this improvement suggests that performance scaling on Tenstorrent hardware is primarily limited by software configuration rather than architectural constraints. Extending similar kernel-level optimization strategies to additional dominant layers is projected to yield an overall 8--12$\times$ cost-normalized improvement relative to the L40S baseline, compared to the current 3.6$\times$ system-level gain.

\subsection{Compute Reduction}

Co-optimization enabled substantial arithmetic reduction:

\begin{itemize}
    \item 4$\times$ compute reduction in the diffusion acoustic model
    \item 8$\times$ compute reduction in the neural vocoder
\end{itemize}

These reductions were achieved through selective low-fidelity execution and block floating-point deployment.

\subsection{Memory Traffic Reduction}

Memory efficiency improvements include:

\begin{itemize}
    \item 2$\times$ reduction in model size
    \item 1.8$\times$ reduction in memory transfer volume
\end{itemize}

These gains were amplified by on-chip data reuse and Network-on-Chip multicast capabilities, reducing redundant DRAM accesses.

\section{Discussion}

\subsection{Limitations}

Despite the reported gains, several limitations remain.

\paragraph{Precision Sensitivity.}
Certain layers exhibit high numerical sensitivity and cannot yet be executed under reduced-fidelity or block floating-point formats without perceptual degradation. This limits full-model low-precision coverage.

\paragraph{Compiler Maturity.}
Program configurations are not fully optimized. Kernel scheduling, memory tiling, and data movement patterns remain areas for improvement. Observed layer-level headroom suggests that current performance does not yet saturate architectural limits.

\subsection{Future Work}

Future efforts will focus on extending hardware–software co-optimization across the full inference graph.

Layer-level measurements indicate that systematic kernel specialization could yield an overall 8--12$\times$ cost-normalized improvement relative to the L40S baseline, compared to the current 3.6$\times$ deployment gain.



We also plan to deploy Lightning V3.1 on Tenstorrent hardware using the same co-design methodology. Lightning V3.1 introduces architectural improvements over V2 and may further increase achievable efficiency.

\subsection{BlockFloat8 Economics}

Native BlockFloat8 support is available in modern high-end GPU architectures; however, such capability is typically confined to premium accelerator tiers. Contemporary GPUs with native BFP8 support are positioned in the approximate \$40,000 price class per device.

In contrast, Tenstorrent enables efficient BlockFloat8 execution on hardware in the approximately \$1,000 price class. This represents an order-of-magnitude difference in hardware acquisition cost for enabling low-precision compute.

The economic implication is that low-precision arithmetic becomes broadly deployable rather than restricted to premium infrastructure. For real-time TTS systems—particularly mid-sized models where accelerator cost dominates deployment decisions—this materially alters the cost-performance tradeoff.

\subsection{Structural Implications}

The results presented here are not solely a consequence of numerical quantization, but of coordinated hardware–software co-design:

\begin{itemize}
    \item Network-on-Chip multicast reduced redundant memory transfers.
    \item SRAM-local tiling reduced DRAM bandwidth pressure.
    \item Selective BlockFloat8 execution reduced arithmetic cost.
\end{itemize}

Together, these factors produced a 4$\times$ cost reduction without perceptual quality loss.

Reducing inference cost at this magnitude expands deployment feasibility for on-prem and latency-sensitive applications, particularly in environments where high-end accelerator budgets are prohibitive.

The primary contribution of this work is therefore not only performance improvement, but a demonstration that TTS systems can be economically optimized through precision-aware hardware co-design without sacrificing perceptual fidelity.

\section{Conclusion}

Diffusion-based text-to-speech systems operate in continuous signal space and exhibit significantly higher numerical sensitivity than token-based language models. Small rounding perturbations can propagate across denoising steps and manifest as perceptible waveform artifacts. As a result, aggressive precision reduction in TTS requires careful, end-to-end validation rather than reliance on conventional tensor-level similarity metrics.

In this work, we demonstrated that precision reduction is nonetheless feasible when guided by empirical sensitivity analysis and hardware-aware execution strategies. Lightning V2 was deployed with approximately 95\% of operations executing under reduced computational fidelity (LoFi) and roughly 80\% of layers utilizing BFP8, while preserving perceptual audio quality.

Through coordinated hardware–software co-design—including selective low-precision execution, SRAM-aware tiling, and Network-on-Chip data movement optimization—we achieved a 4$\times$ reduction in accelerator cost relative to an NVIDIA L40S baseline for equivalent workload capacity.

These results indicate that inference efficiency in TTS is not constrained solely by model architecture, but by how numerical precision, memory movement, and hardware scheduling interact. 

This work demonstrates that inference economics can be reshaped through precision-aware model design and hardware co-optimization.

\bibliographystyle{unsrt}  
\bibliography{references}

\end{document}